\documentclass{vgtc}                          

\ifpdf
  \pdfoutput=1\relax                   
  \usepackage{graphicx}                
  \DeclareGraphicsExtensions{.pdf,.png,.jpg,.jpeg} 
\else
  \usepackage{graphicx}                
  \DeclareGraphicsExtensions{.eps}     
\fi%

\graphicspath{{figures/}{pictures/}{images/}{./}} 

\usepackage{microtype}                 
\PassOptionsToPackage{warn}{textcomp}  
\usepackage{textcomp}                  
\usepackage{mathptmx}                  
\usepackage{times}                     
\usepackage{cite}                      
\usepackage{tabu}                      
\usepackage{booktabs}                  

\usepackage{amsfonts}
\usepackage{subfigure}

\onlineid{0}

\vgtccategory{Research}

\vgtcinsertpkg

\title{Conditional Parallel Coordinates}

\author{Daniel Karl I. Weidele\thanks{e-mail: daniel.karl@ibm.com}\\ %
        \scriptsize IBM Research AI%
    }

\abstract{Parallel Coordinates \cite{inselberg1985plane, inselberg1990parallel} are a popular data visualization technique for multivariate data. Dating back to as early as 1880\cite{gannett1883general} PC are nearly as old as John Snow's famous cholera outbreak map \cite{snow1855mode} of 1855, which is  frequently regarded as a historic landmark for modern data visualization. Numerous extensions have been proposed to address integrity, scalability and readability. We make a new case to employ PC on conditional data, where additional dimensions are only unfolded if certain criteria are met in an observation. Compared to standard PC which operate on a flat set of dimensions the ontology of our input to Conditional Parallel Coordinates is of hierarchical nature. We therefore briefly review related work around hierarchical PC using aggregation or nesting techniques. Our contribution is a visualization to seamlessly adapt PC for conditional data under preservation of intuitive interaction patterns to select or highlight polylines. We conclude with intuitions on how to operate CPC on two data sets: an AutoML hyperparameter search log, and session results from a conversational agent.
} 

\CCScatlist{
  \CCScatTwelve{Human-centered computing}{Visu\-al\-iza\-tion}{Visu\-al\-iza\-tion techniques}{};
  \CCScatTwelve{Human-centered computing}{Visu\-al\-iza\-tion}{Visualization design and evaluation methods}{}
}

\begin{document}


\firstsection{Introduction}

\maketitle

\begin{figure}[h]
	\centering
	\includegraphics[width=0.9\linewidth]{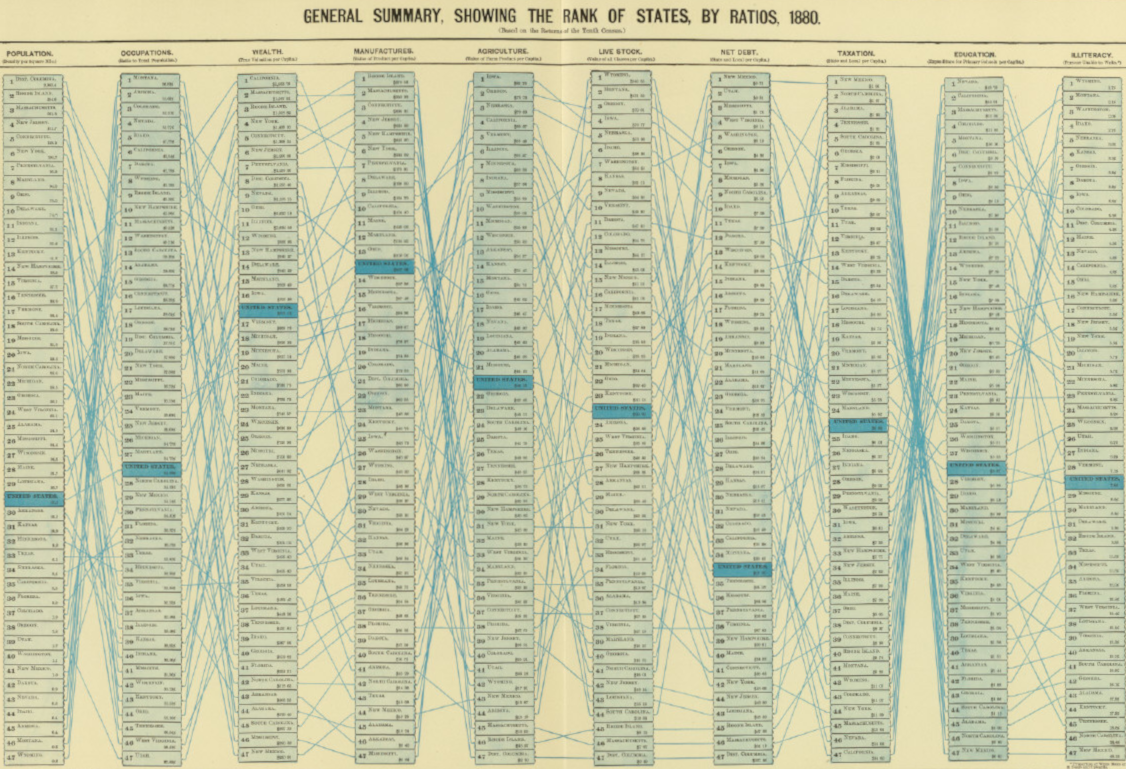}
	\caption{Early PC visualization from 1880 \cite{gannett1883general} showing the United States ranked by population, wealth, live stock, net debt, and others. The blue highlighted observation represents the U.S. average.}
	\label{fig:first}
\end{figure}

Parallel Coordinates (PC) are a fundamental technique to visualize multivariate data. At least dating back to 1880 \cite{gannett1883general} (Figure \ref{fig:first}) the visualization method is nearly as antique as John Snow's famous Cholera outbreak map of 1855 \cite{snow1855mode}, which is often highlighted as an early landmark of modern data visualization. Given an additional renaissance \cite{inselberg1985plane, inselberg1990parallel}, simple applicability, and their genericity to adapt different data types make PC a well-known exploratory component in the data scientist's toolbox.

Multivariate data are multi-dimensional observations with values typically drawn from different scales and types. For example, we can observe cars as 3-dimensional data containing license plate (nominal scale), number of doors (discrete scale) and date of registration (interval scale). Clearly such observations are underlying many experimental research setups, so it is tempting to claim multivariate data is everywhere.

In its most basic form PC visualizations map each dimension to a vertical axis, and then draw each observation as a horizontal polyline. Lines intersect axes at where the data value can be found on the axis. For example, an observed value of 0.5 would make the line cross a numerical axis with minimum 0, maximum 1 exactly in the middle. Nominal values can be treated accordingly, ordered depending on the application.

The contribution of this paper is three fold. First, we introduce \emph{conditional data} as a novel arrangement of multivariate data. Second, we adapt PC for conditional data under preservation of intuitive interaction patterns. Third, we provide intuitions on how to operate CPC on two data sets: an AutoML hyperparameter search log, and session results from a conversational agent.

\section{Method}
Multivariate data can be conditioned on properties of the observation: one might need a set of dimensions appropriate for a product of type pizza - toppings and diameter, for example, and one for soft drinks - size and flavor. Yet, both products share delivery instructions and payment options. In the following we provide a formal characterization of such conditional data, and introduce CPC as a novel technique for their visualization.

\subsection{Conditional Data}
Consider a set of predicates $C := \{C_1, ..., C_k\}$ on $d$-dimensional observations $o \in O$, where $O := \{D_1 \times ... \times D_d\}$ for some multivariate $D_j \in \{\mathbb{N}, \mathbb{R}, \{0, 1\}\}$ and $j \in [1, d]$:

\begin{equation}
	C_i: O \rightarrow \{true, false\} \textrm{ , with } i \in [1, k].
\end{equation}
Let further $O_{C_i}$ be the subset of observations for which $C_i$ holds:

\begin{equation}
O_{C_i} := \{o \in O~|~C_i(o) = true\}.
\end{equation}
Then \emph{conditional data} is the union of tuples $(o_i, \hat{o}_i)$ where $o_i \in O_{C_i}$ and $\hat{o_i} \in \hat{O_i}, \hat{O_i} := \prod\limits_{l=1}^{e_i} D_l$ some additional $e_i$-dimensional multivariate observation:

\begin{equation}
\mathcal{C} := \bigcup\limits_{i=1}^{k} (o_i, \hat{o}_i).
\end{equation}
In other words, if a criteria is met for an observation we are allowed to augment it with further details. Yet, such additional information will not even \emph{exist} in cases where the predicate does not hold (i.e. no toppings on soft drinks). We can further recurse the process by setting $O = \mathcal{C}$ and defining more predicates.

For this work we consider simple predicates limited to single variables, thus conceptually binding the additional information to a specific value (or range) of that variable. This reduces items of $\mathcal{C}$ to tree-like structures without well-defined, single roots, yet additional information branching out when the predicate is met on a variable.

\subsection{Extending Parallel Coordinates}
\begin{figure}[h]
	\centering
	\includegraphics[width=0.48\linewidth]{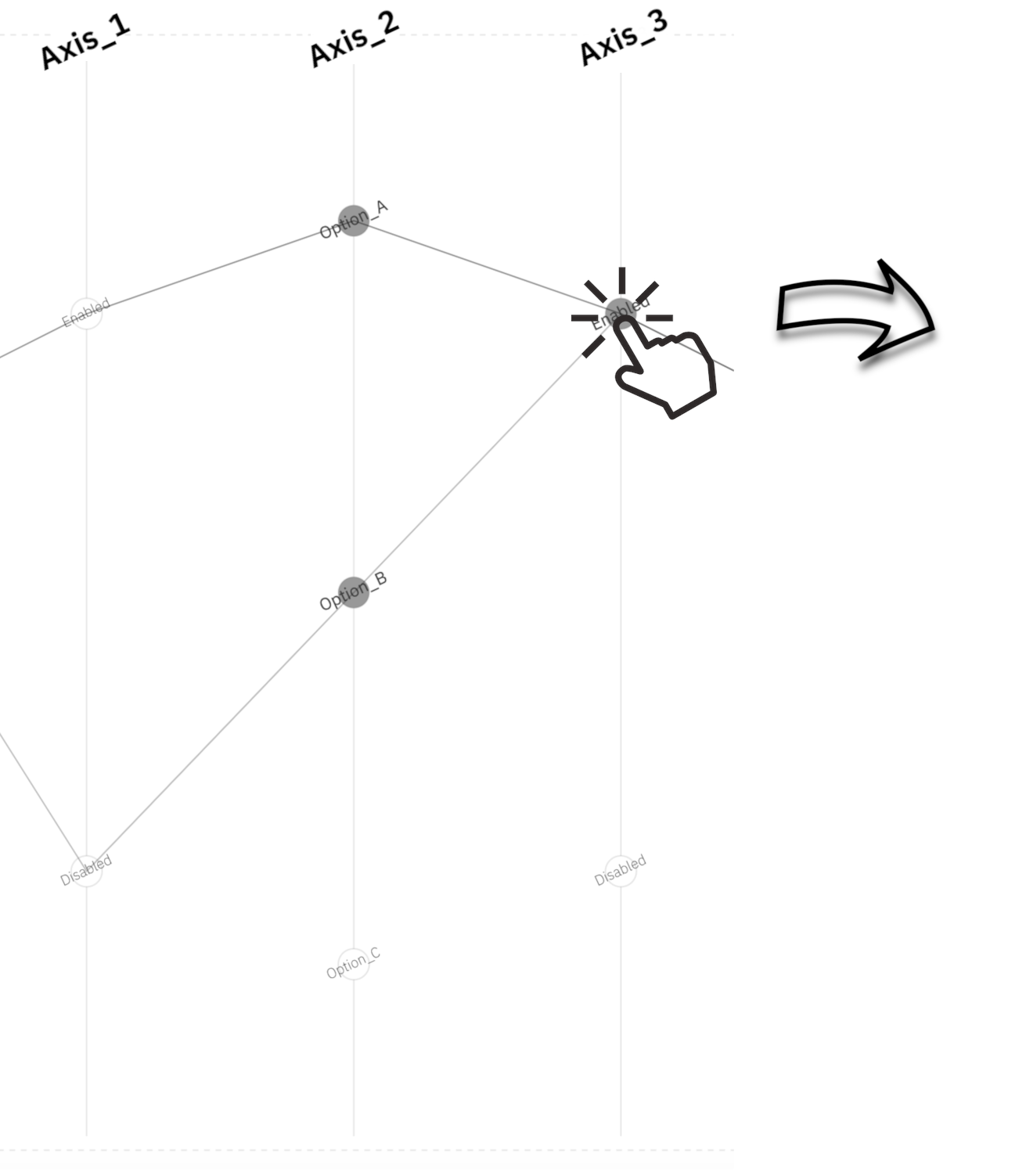}
	\includegraphics[width=0.48\linewidth]{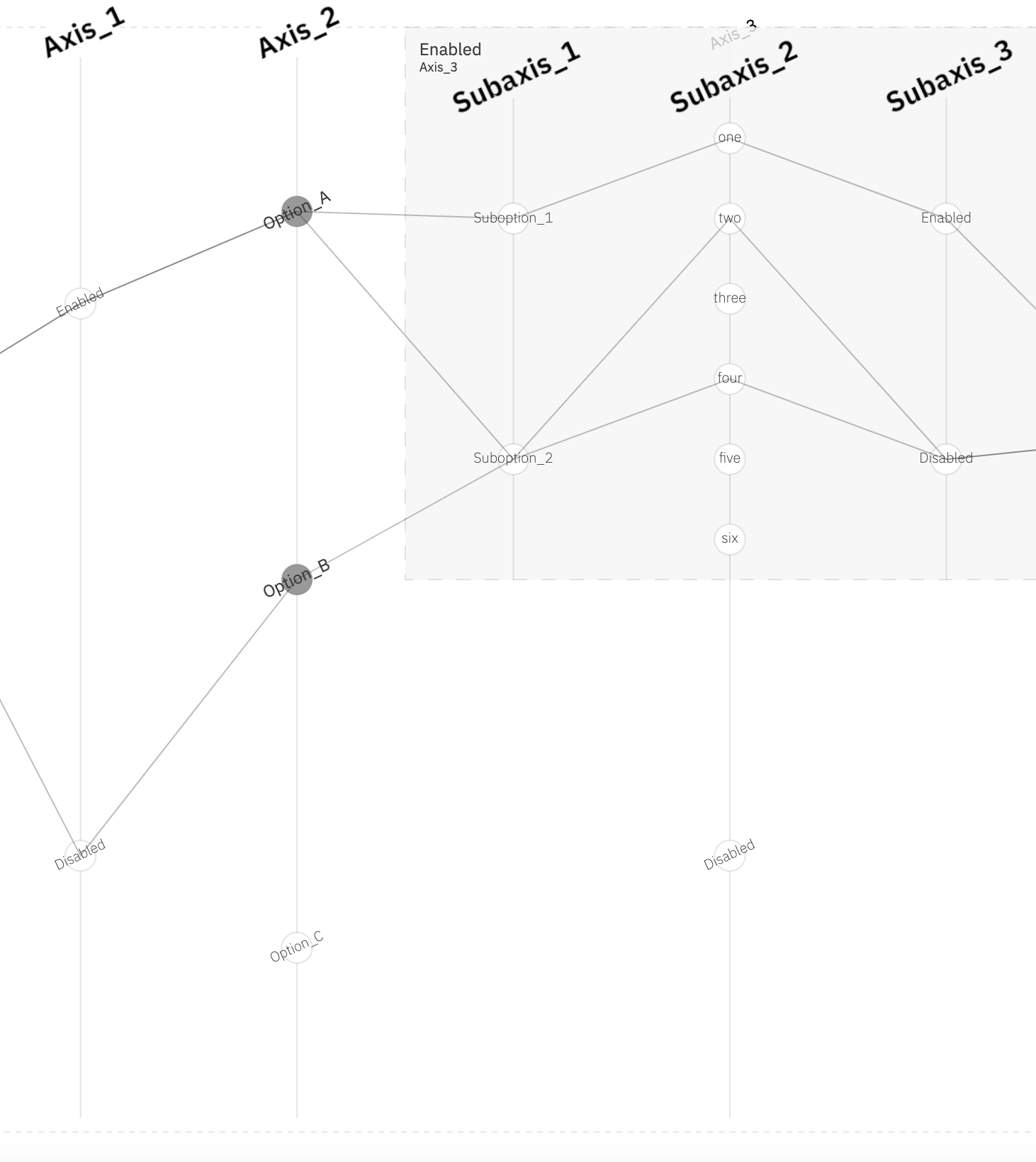}
	\caption{Illustration of interaction with Conditional Parallel Coordinates. Since a predicate has been defined and additional information is available for the \emph{Enabled} option of Axis 3. Such expandable options (gray) can be clicked to seamlessly expand sub-dimensions. As we get to see more details the expansion step reveals there have in fact been 3 polylines. In the example the user can further expand Options A and B on Axis 2. The result is depicted in \autoref{fig:cpc_exp}.}
	\label{fig:cpc}
\end{figure}

The above characterization of conditional data for simple predicates almost naturally reveals a variant of parallel coordinates. With additional information $\hat{o}_i$ being bound to a particular value (range) we can visually unfold $\hat{o}_i$ when a user clicks (brushes) the corresponding visual artifact. Figure \ref{fig:cpc} shows the expansion process after the user selected the upper value of Axis 3.




For categorical variables we initially space out the options equally, offsetting them by half a height. Then, the canvas for $\hat{o}_i$ can be placed centered on top of an option (or a selected range), such that no overlap would be created on the y-axis upon expansion.
To accommodate the increased space requirement on the x-axis we sum a weight $w$ over all visible dimensions, recursively stepping into branched options. In the predicate-free or collapsed trivial case we return $w$, which we set to $1$. For dimensions with multiple expanded branches we return the maximum. Depending on the application $w$ can be chosen more fine-grained.

\begin{figure}[h]
	\centering
	\includegraphics[width=0.8\linewidth]{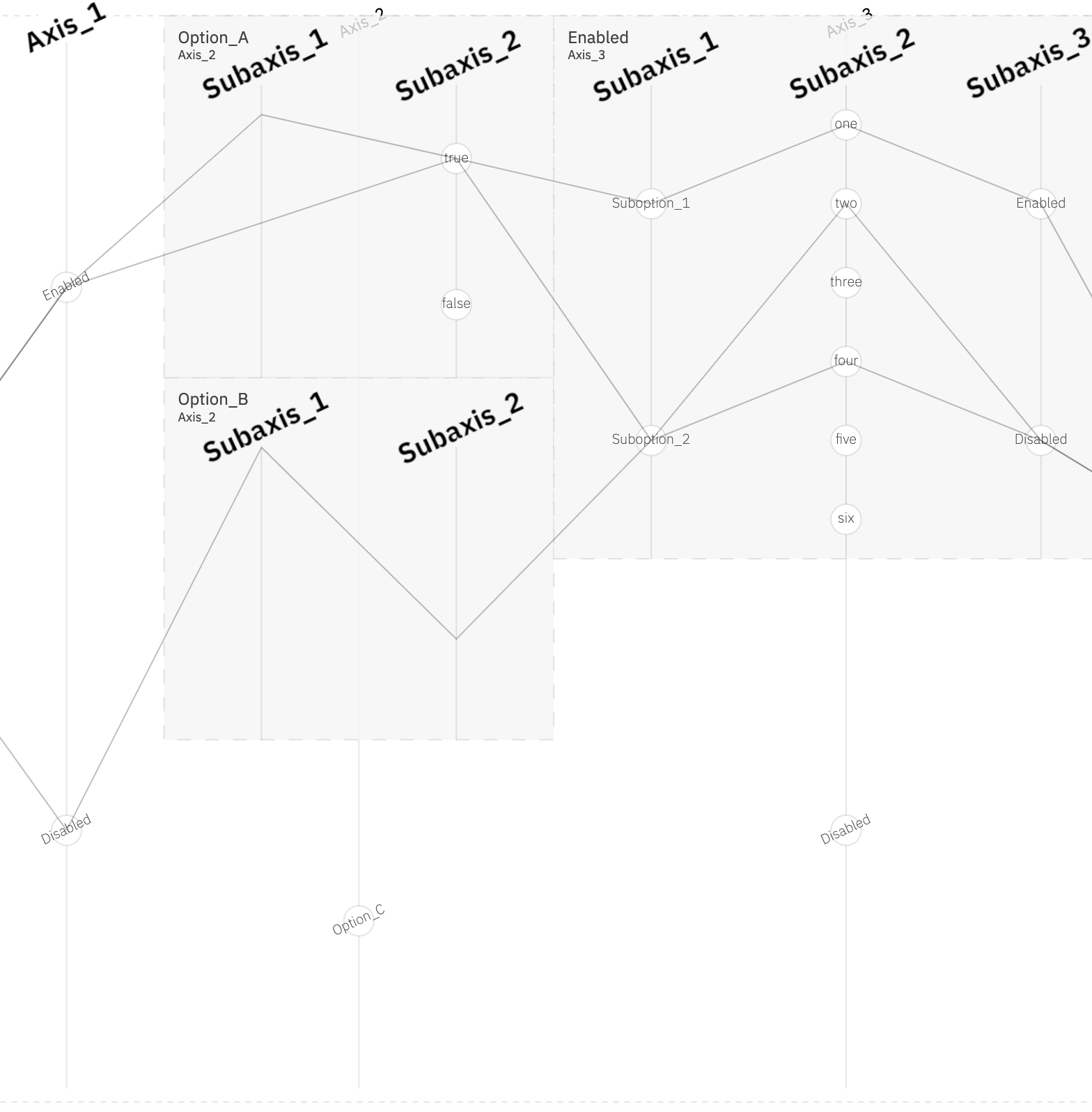}
	\caption{Status of the visualization after all hierarchical options have been expanded: The component reserves more room horizontally and vertically for  collapsible options, such that sub-axes can be inserted without causing any overlap. The technique can further be applied recursively (collapsible options within sub-axes), or when brushing ranges.}
	\label{fig:cpc_exp}
\end{figure}

Figure \ref{fig:cpc_exp} represents the state of the visualization after all additional data has been expanded. In our current conceptual prototype we do not handle over-plotting of equivalent $o_i$ in the top level, however, the additional dimensions allow us to separate observations based on their more fine-grained sub-dimensions. Furthermore, since the fully collapsed state of CPC is equivalent to PC, we can apply proposed variants of PC as well, such as visualizing probabilities \cite{unwin2003parallel, dasgupta2010pargnostics}. Moreover, not breaking the basic principles of PC in the visualization of sub-dimensions allows to transfer and apply such variants in many cases, too. In this regard, \autoref{sec:extensions} outlines how common polyline highlighting can be intuitively realized in CPC.

In the next section we will review and discuss related work around PC, before looking at extensions and applications of CPC.

\section{Related work}
\label{sec:related_work}

To the best of our knowledge the proposed extension of PC is novel. We conducted a thorough literature review on PC and briefly present our findings comparing CPC to related variants. Andrews et al. \cite{andrews2015aggregated} integrate hierarchical dimensions into parallel coordinates. Compared to CPC, however, additional dimensions are tied to a dimension as a whole rather than particular values (or ranges) of the dimension. The data further is not necessarily hierarchical, rather dimensions on higher levels \emph{hold place}, e.g. substituting a number of dimensions with the mean. This substitution is also reflected in the visualization, where overview and detail axes are not displayed simultaneously. In \cite{artero2004uncovering} Artero et al. aim to reveal clusters in crowded parallel coordinates. The authors aim to uncover stronger signals among large numbers of observations by directly computing pixel intensities for polylines. Similarly Fua et al. \cite{fua1999hierarchical} cope with PC for large data sets by computing a hierarchy for polylines. Then, at different scales lines are replaced by proxies to show more and more fine-grained trends in the data. Angular brushing \cite{hauser2002angular} is an extension for PC that allows users to define angles on axes. Polylines whose slope fall in the given angular range are then being selected for further processing. The technique has not been particularly proposed for sub-dimensions, however, it confirms the advantage of not breaking basic principles of PC since it is straight-forward to apply angular brushing in CPC, too.

\begin{figure}[h]
	\centering
	\includegraphics[width=\linewidth]{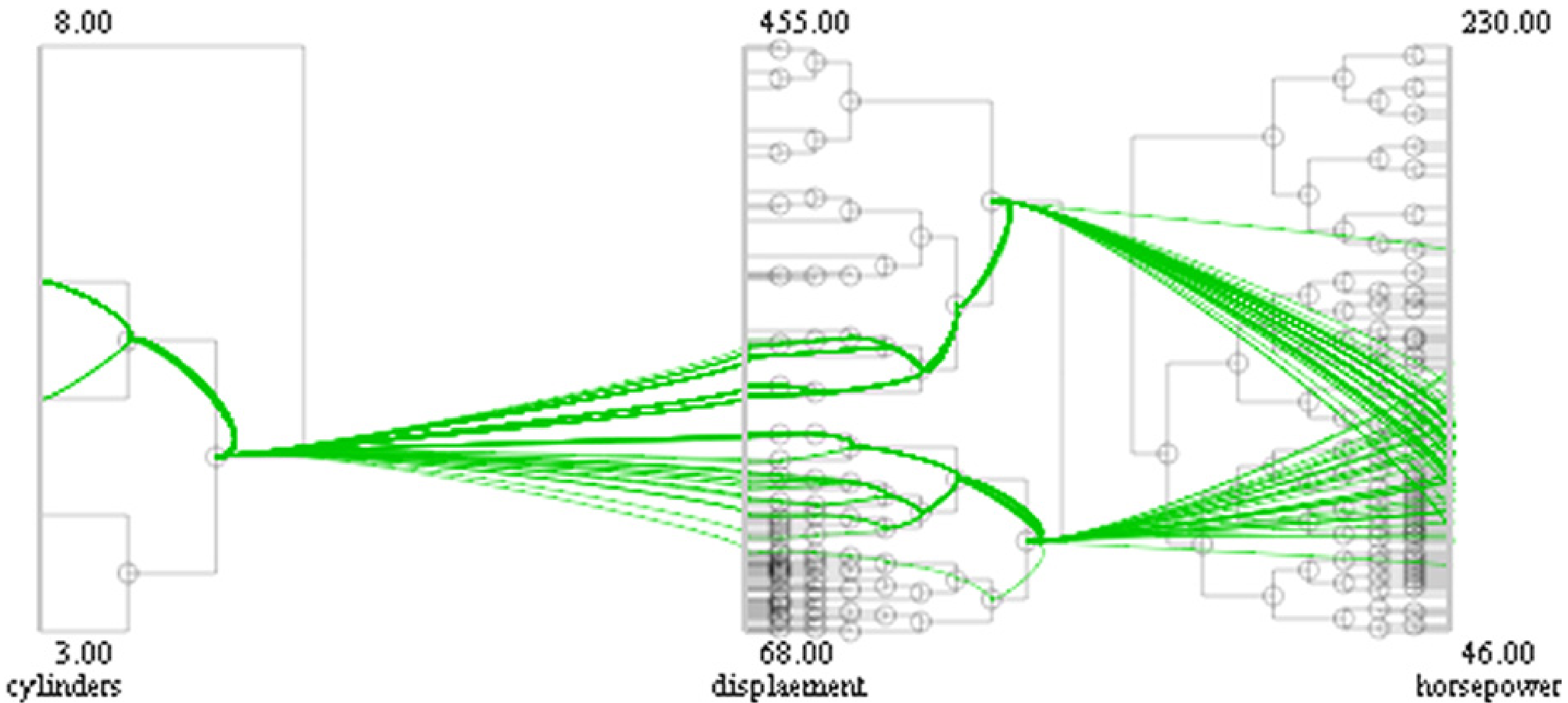}
	\caption{Dendrograms attached to axes of PC \cite{huang2016novel}}
	\label{fig:ctree}
\end{figure}

We believe a visualization given in Huang et al. \cite{huang2016novel} is most similar to our proposed method (see Figure \ref{fig:ctree}). Here, the authors attach dendrograms to axes of parallel coordinates. While not having been proposed as such, we can imagine to visualize conditional data rather than clustering information in the form of trees in this way. Then, we find our more tree\emph{map}-like approach \cite{shneiderman1992acm} has some advantages. While not having proposed an interactive mechanism, mixing trees and PC in their proposed way imposes additional complexity as the user has to decode two data structures simultaneously. Since different values of sub-dimensions are further not drawn at the same position on the x-axis it is hard to compare polylines. We suppose staying within the known makes CPC favorable in handling different data types, where trees would require binning. Lastly---and probably most importantly---tree nodes can only hold values of \emph{one} dimension and then immediately split into child nodes. That means additional dimension information could not be shown on the same level. Therefore, encoding with trees cannot capture the full expressiveness of conditional data. Parallel Hierachies \cite{vosough2018parallel} is a similar technique to \cite{huang2016novel}, where additional sub-categeories are displayed more interactively instead of showing the tree as a whole directly. Novotny and Hauser \cite{novotny2006outlier} combine PC with a matrix visualization capturing trajectories between any 2 neighboring axes. The axes are mapped to columns and rows of the matrix, and cells are colored based on numbers of connecting polylines. The matrix can support to reveal patterns in the data. Clearly such matrices can be shown between sub-dimensions in CPC, too. The Focus+Context representation proposed by Richer et al. \cite{richer2018enabling} organizes and aggregates axes values in blocks, and visually encodes underlying distributions in form of histograms. When the user drills into blocks, focus is shifted towards details of the selected distribution. The approach appears to provide a more multi-scalar perspective on single dimensions, working against different densities in the data.  Wang et al. \cite{wang2016multi} address \emph{ensembles} of polylines. Based on two alternatives to represent the groups the authors derive a hybrid approach combining the advantages of both. A smooth parameterization then allows to transition between \emph{superimpostion}, where polylines sets are represented in a single PC space, and \emph{juxtaposition}, where the groups are  plotted on individual axes per group, and then stacked.

\section{Adapting interaction patterns}
\label{sec:extensions}

In this section we briefly discuss how \emph{polyline highlighting} can be adapted for CPC. Moreover we present an \emph{edit mode} for CPC, to manually draw observation data. 

\subsection{Highlighting}

Polyline highlighting is an effective interaction mechanism in Parallel Coordinates. Typically, hovering or selecting a line visually emphasizes it to help follow through the observation across all dimensions. Other practices for highlighting in PC include hovering/selecting values on categorical axes, or brushing ranges on numerical axes. Then, all lines intersecting with the value or falling inside the range are accentuated. In this way multiple lines can be put into focus simultaneously, enabling the user to more easily spot trends or identify patterns of correlation---especially when correlating axes are not placed next to each other.

\begin{figure}[h]
	\centering
	\includegraphics[width=\linewidth]{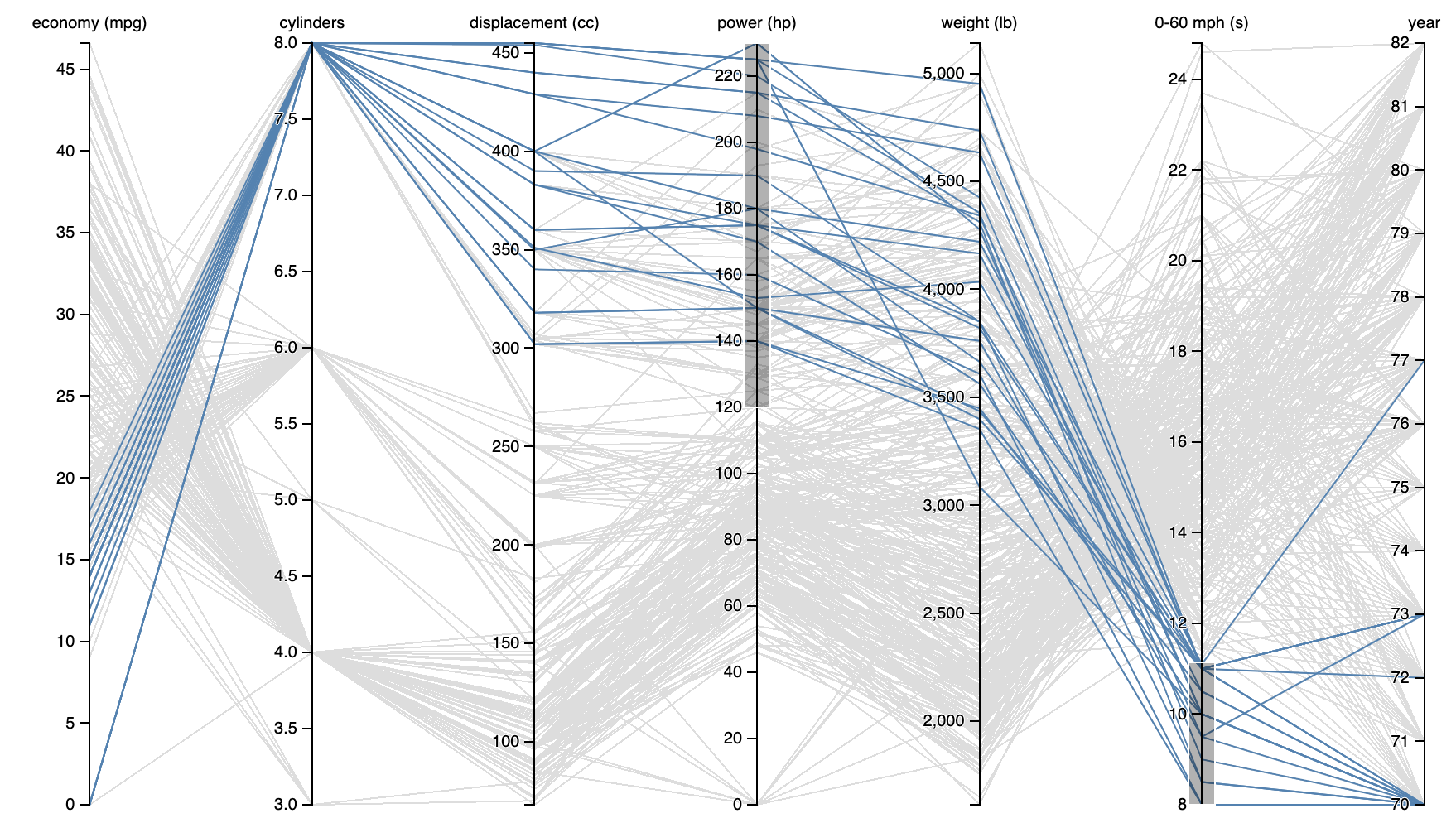}
	\caption{Typical highlighting in classic Parallel Coordinates. When hovering an option (nominal attributes) all polylines crossing the option are highlighted. In case of numerical attributes users can brush ranges to highlight polylines. The example configuration shows most fast cars of data sets are 8-cylinder muscle cars from the early 1970s\textsuperscript{\ref{ftn:classic_highlighting}}.}
	\label{fig:classic_highlighting}
\end{figure}

Figure \ref{fig:classic_highlighting} shows a range brushing example for classic PC in the D3 visualization library\footnote{Jason Davies. Parallel Coordinates. In \url{https://bl.ocks.org/jasondavies/1341281}, accessed on June 10, 2019.\label{ftn:classic_highlighting}}. The underlying observations are cars for which typical criteria have been measured: gas mileage, weight, power, etc. We filter for higher power cars taking less time to accelerate from 0 to 60 $\frac{miles}{h}$. 
Polyline highlighting then immediately reveals in the data this criteria mostly holds for 8 cylinder muscle cars from the early 1970s.

To support highlighting of polylines in CPC when dealing with additional sub-dimensions we propose the following mechanisms. Since polylines are not broken when entering sub-dimensions we can implement line hovering the same way as in CP. For value/range highlighting, however, we would lose the visual elements indicating values on axes in the expanded state, when sub-dimensions are visible. We therefore render bounding boxes around additional dimensions. Then, hovering the background of bounding boxes has the same effect as hovering non-expanded values, such that all polylines cutting a hovered box are emphasized.

\begin{figure}[h]
	\centering
	\subfigure[Highlighting via bounding box of expanded option]{%
		\includegraphics[width=.4\linewidth]{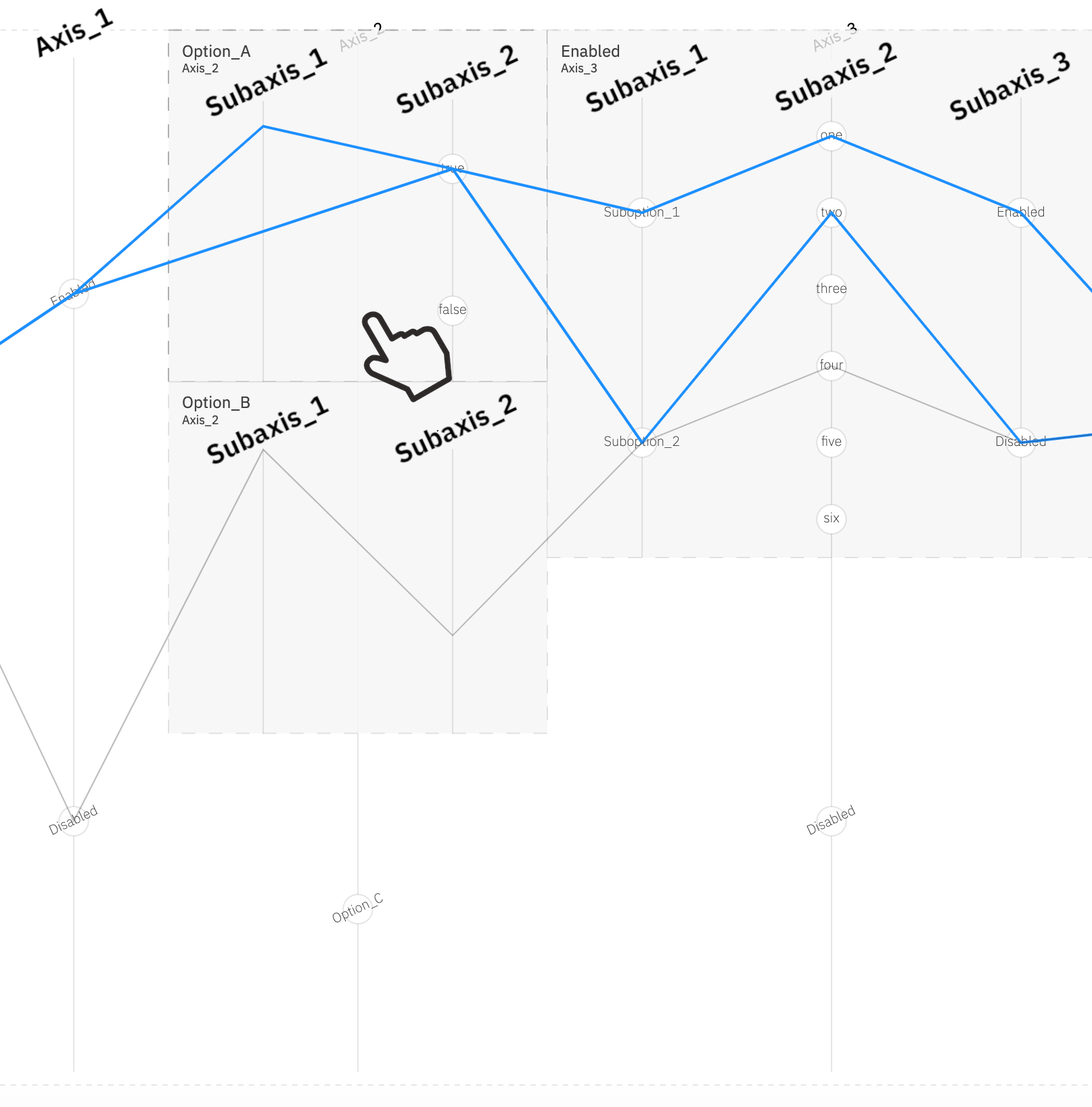}
		\label{fig:cpc_highlighting_1} 
	} 
	\quad 
	\subfigure[Highlighting through option in sub-dimension]{%
		\includegraphics[width=.4\linewidth]{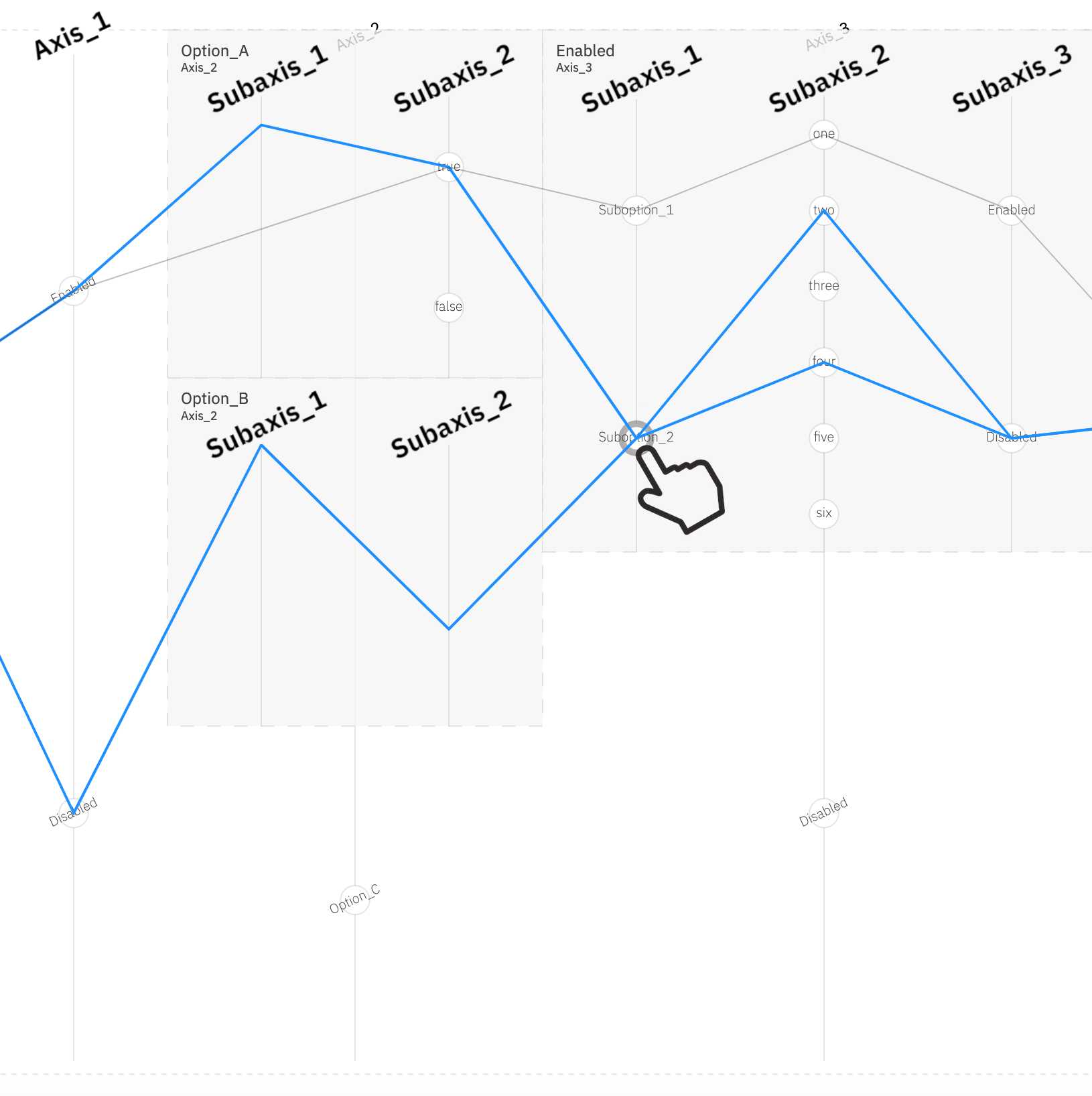}
		\label{fig:cpc_highlighting_2} 
	} 
	\caption{Polyline highlighting in Conditional Parallel Coordinates}
	\label{fig:cpc_highlighting}
\end{figure}

Figure \ref{fig:cpc_highlighting} shows the alternative highlighting scenarios implemented in our prototype. Hovering the bounding box for Option A on Axis 2 highlights the two upper polylines (\autoref{fig:cpc_highlighting_1}), whereas pointing on Suboption\_2 of Subaxis\_1 in the Enabled value of Axis\_3 accentuates the upper and lower polyline.



\subsection{Edit mode}

We present another mode of operation for CPC that allows users to manually draw additional observations. In this \emph{edit mode} the user constructs a polyline by sequentially picking values from the axes. Instead of drawing whole polylines from scratch, the user can also duplicate existing polylines to then adjust only some values of the copy. This can be useful in the data collection process, to quickly generate more samples, or to provide visual analytics feedback loops as presented in \autoref{sec:automl}, where the edit mode speeds up the process feeding hand-drawn special cases back into the system. 

\begin{figure}[h]
	\centering
	\includegraphics[width=0.8\linewidth]{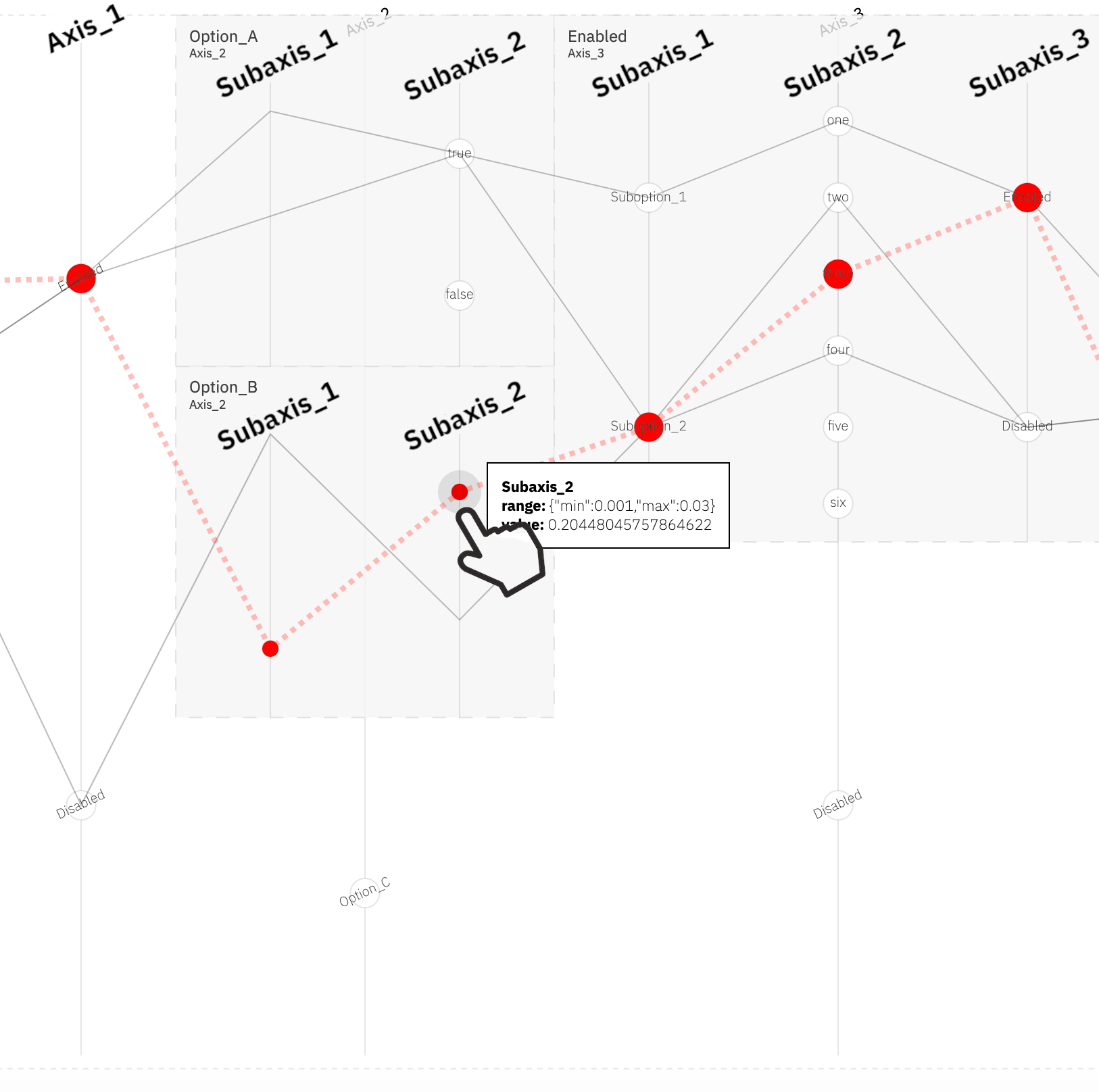}
	\caption{Edit mode: Users can record custom polylines (red) by manually drawing them directly on top of the CPC visualization.}
\end{figure}

For the implementation of edit mode we find the following criteria noteworthy to share:
(1) Selecting a value from sub-dimensions $\hat{o}_i$ induced by predicate $C_i$ removes selections in any other sub-dimensions $\hat{o}_j$ for all $C_j$  defined on the same dimension as $C_i$.
(2) Moreover, selecting a value from sub-dimensions removes a potentially existing direct selection in an alternative path of the upper dimension, and vice versa.
(3) When in edit mode we disable highlighting, show additional tooltips and keep the edited polyline focused in an alternative color.

\section{Applications}
We provide two exemplary use cases of CPC. The first is a direct application of our prototype to log data of a software that automatically builds a machine learning pipeline and fine tunes its hyperparameters. Secondly, we describe a use case of CPC on session results from conversational agents.

\subsection{Hyperparameter Search in AutoML}
\label{sec:automl}

\begin{figure}[h]
	\centering
	\subfigure[Fully collapsed pipelines reducing CPC to classic PC]{%
		\includegraphics[width=.9\linewidth]{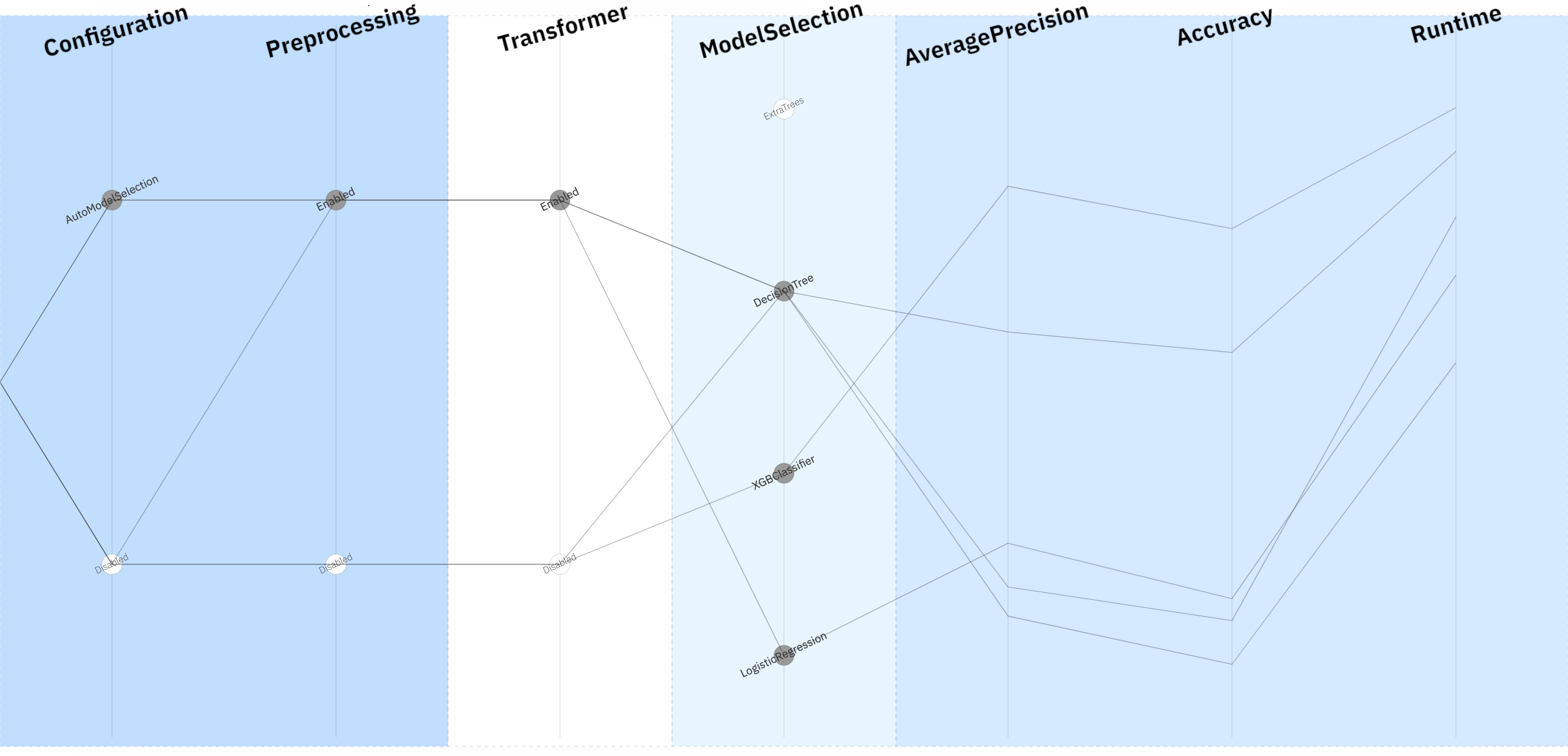}
		\label{fig:automl_collapsed} 
	} 
	\\
	\subfigure[Fully expanded pipelines with additional hyperparameter settings]{%
		\includegraphics[width=.9\linewidth]{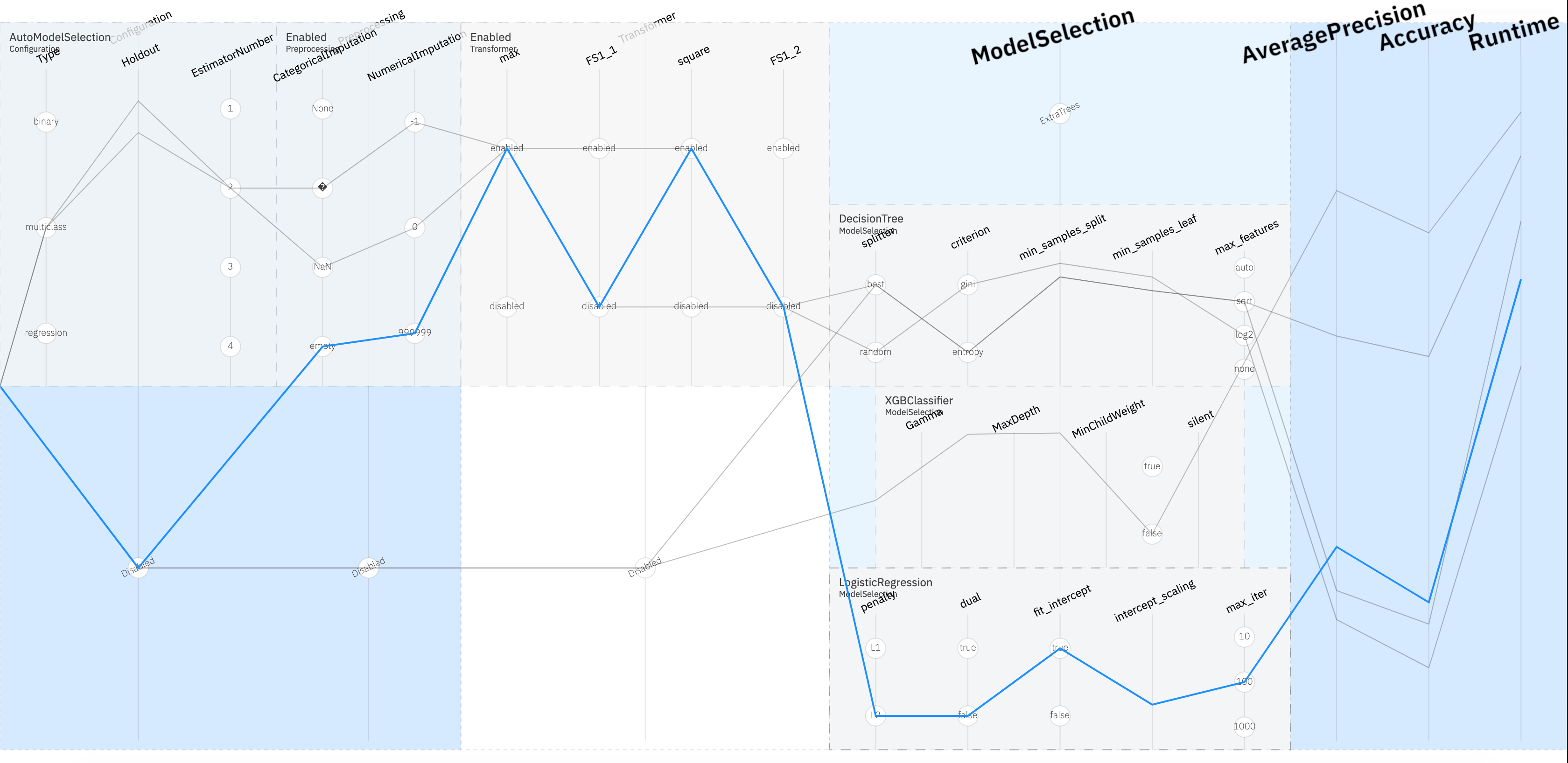}
		\label{fig:automl_expanded} 
	} 
	\caption{CPC applied to AutoML search log. Tested pipelines are mapped to polylines of the visualization. Additional hyperparameter information can be obtained through expansion of a configuration.}
	\label{fig:automl}
\end{figure}

Parallel Coordinates have recently been applied in visualizations of automatic machine learning software \cite{wang2019atmseer, parkvisualhypertuner}. AutoML typically assembles individual algorithmic blocks into a pipeline, and then fine-tunes hyperparameters of the blocks. PC is not capable of showing the pipeline and hyperparameters simultaneously. We therefore apply CPC to the pipeline rendering steps sequentially from left to right, and require predicates to be of form \emph{equal to block ID}. If a pipeline matches the predicate we augment the block with corresponding sub-dimensions showing its hyperparameter settings. \autoref{fig:automl_collapsed} depicts 5 tested pipelines in our prototype, with all blocks collapsed. After expanding all blocks \autoref{fig:automl_expanded} shows details about hyperparameters while preserving the context. In this example the last 3 axes are performance measurements, so the user can easily hover a high accuracy pipeline and quickly learn about the configuration. If AutoML has not tested certain configurations (i.e. when stuck in local minima), the user can leverage the \emph{edit mode} to close the feedback loop and inform the search algorithm about a new initial configurations.

\subsection{Session Results of Conversational Agent}
\label{sec:chatbot}
Conversational agents \cite{weizenbaum1966eliza, allen1995trains, abdul2015survey} become more and more commonplace continuing to replace real world service help desks. Companies accumulate transcripts of chat sessions to further improve the customer experience. We see a chance to gain new insights into such data by exploring it through the lens of CPC, especially deeper CPC with further recursions.
Consider a pizza booth chat bot with the following data mapped to main axes of CPC: a number of axes for food items (with options soft drink, pizza, salad, dessert, pasta), an axis for delivery type (self-pickup, eat-in, delivery) and an axis for the payment option (cash, credit card, online payment solutions, etc.).
We can augment food item selections with additional information via sub-dimensions, i.e. pizzas have toppings and a diameter, whereas soft drinks have  sizes on a different scale, brands, and flavor. Salads in turn have dressings, and different bases like lettuce, kale or chicory as sub-dimensions. While self-pickup or eat-in options might not carry a lot of additional information, the delivery option sure will require a city, and zip code. Lastly the credit card or online payment option could be augmented with data from royalty programs. Exploring the chat session results as polylines of a CPC visualization can inform the user about trends in the data, especially when further filtering for polylines from a certain day, week or month to explore seasonalities in different granularities.

\subsection{Discussion}
Our proposed extensions are especially useful in cases when limiting focus to particular examples. Expansion of options reduces consumed screen space, while highlighting preserves the ability to follow through selected polylines end-to-end. In cases where many options are expanded, however, the visualization might introduce new uncertainty as polylines can cross within axis. This could be further improved by edge bundling \cite{zhou2013edge, palmas2014edge} or edge routing \cite{dobkin1997implementing}. The edit mode has mostly been motivated by our application \autoref{sec:automl}, and proven useful for an interactive feedback loop.

\section{Conclusion and Future work}
We characterized conditional data, i.e. data that can be recursively augmented with more details if certain criteria is met. In the case of simple predicates only defined on single values or ranges we propose Conditional Parallel Coordinates as a novel extension of Parallel Coordinates. Items for future work are (1) visualization techniques to address more complex predicates, (2) parallelism in PC/CPC if observations do not have full overlap in dimensions (likely the case for range selections on numeric dimensions), and (3) exploring the concept of conditional data towards \emph{smarter}, more functional data in general---favorably building around the theory of relational algebra.

\acknowledgments{
The author would like to thank Gregory M. Bramble, Horst Samulowitz and Dakuo Wang for the provision of the AutoML data set, as well as Parikshit Ram, Hendrik Strobelt, Steven I. Ross and Mauro Martino for discussions.}

\bibliographystyle{abbrv-doi}

\bibliography{template}
\end{document}